\documentclass[pra,aps,showpacs,tightenlines]{revtex4}
\usepackage{epsfig}
\usepackage{amssymb}
\begin{document}
\title{Large two atom two photon vacuum Rabi oscillations in a high quality
cavity}
\author{P. K. Pathak, and G. S. Agarwal}
\address{Physical Research Laboratory, Navrangpura, Ahmedabad 380 009}
\date{\today}
\begin{abstract}
We predict large cooperative effect involving two atom two photon
vacuum Rabi oscillations in a high quality cavity. The two photon
emission occurs as a result of simultaneous de-excitation of both
atoms with two photon resonance condition
$\omega_1+\omega_2\approx \omega_a+\omega_b$, where $\omega_1$,
$\omega_2$ are the atomic transition frequencies and $\omega_a$,
$\omega_b$ are the frequencies of the emitted photons. The actual
resonance condition depends on the vacuum Rabi couplings. The
effect can be realized either with identical atoms in a bimodal
cavity or with nonidentical atoms in a single mode cavity.
\end{abstract}
\maketitle
\section{introduction}
 High quality cavities have led to the study of a new
regime of radiation matter interaction {\it viz} the study of
strongly interacting systems. Several new phenomena such as vacuum
Rabi splittings \cite{{1},{2},{3},{4}}, collapse and revival of
Rabi oscillations \cite{5}, trapping states \cite{6}, and systems
like micromasers \cite{{7},{8}} have been studied. More recent
applications on high quality cavities are in the context of
quantum computation \cite{9}. Most of these works concern the
interaction of the individual atoms. Earlier cooperative effects
like optical bistability involving a large number of atoms have
been investigated \cite{10}. A large part of these
 studies
concerns the situations where the atomic transition frequency is
almost equal to the cavity frequency. In this paper we report an
unusual cooperative effect involving two atoms in a nonresonant
cavity. This cooperative effect arises from the simultaneous
de-excitation of two atoms such that the sum of the energies of
emitted photons is equal to the sum of the excitation energies of
the atoms.
 We demonstrate that in a high
quality cavity the two atom two photon resonant effect could be
 large thus opening up the possibility of a variety of
nonlinear {\it i.e.} multi-photon cooperative phenomena in
nonresonant cavities. For this purpose the recent development on
the trapping of atom inside the cavity \cite{kimble} should be
especially useful. We bring out the origin of such large two atom
two photon Rabi oscillations.

\begin{figure}[h]
\begin{center}
\includegraphics[width=3.5in,height=2in]{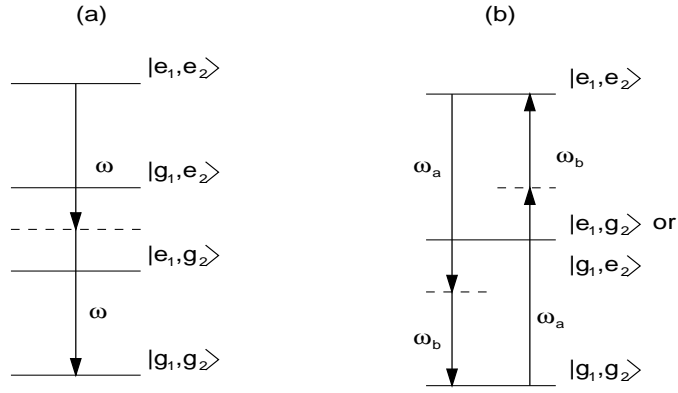}
\end{center}
\caption{Two ways for two atom two photon emission, $(a)$
corresponding to two possible intermediate states
$|e_1,g_2\rangle$ and $|g_1,e_2\rangle$ in the system of
nonidentical atoms interacting with a single mode vacuum, $(b)$ in
the system of identical atoms interacting with two modes of the
vacuum.} \label{fig1}
\end{figure}

We start by noting that in a two photon emission process the two
photon resonance between the excited state $|e\rangle$ and the
ground state $|g\rangle$ would occur at a frequency given by
$\omega_{eg}=2\omega$, where $\omega_{eg}$ is the atomic
transition frequency and $\omega$ is the frequency of the photons
emitted. The process proceeds via intermediate states $|i\rangle$,
which are away from a single photon resonance. Now consider an
inter-atomic process involving two atoms with distinct transition
frequencies $\omega_{1}$ and $\omega_{2}$ such that
$\omega_{1}-\omega$ and $\omega_{2}-\omega$ are large so that
individual emissions are not important. However, as shown in
Fig.\ref{fig1}(a), one can consider a two photon emission process
such that $\omega_{1}+\omega_{2}=2\omega$. Clearly this would be a
cooperative process as it involves two atoms. Besides it should
also be important as it is a resonant process. Let us then examine
the transition probability for such a two photon emission. Let
$H_+$ be the interaction responsible for the emission of a photon
defined by the interaction Hamiltonian which is written in the
form
\begin{equation}
H_I=H_+e^{i\omega t}+H_-e^{-i\omega t}.
\end{equation}
Then the second order perturbation theory leads to the following
expression for the rate of two photon emission
\begin{eqnarray}
R_{c}=\frac{2\pi}{\hbar^2}\left|\frac{\langle
g_1,g_2|H_+|g_1,e_2\rangle\langle
g_1,e_2|H_+|e_1,e_2\rangle}{\hbar(\omega_{1}-\omega)}
+\frac{\langle g_1,g_2| H_+|e_1,g_2\rangle\langle
e_1,g_2|H_+|e_1,e_2\rangle}{\hbar(\omega_{2}-\omega)}\right|^2
\delta(\omega_{1}+\omega_{2}-2\omega).
\end{eqnarray}
Note that surprisingly $R_{c}=0$, as the two photon matrix element
vanishes when $\omega_1+\omega_2=2\omega$ as there are two paths
for two photon emission which interfere destructively. It has been
argued that a nonzero two photon emission can result if we include
inter-atomic interactions \cite{{11},{12}} which, however, are
important only if the inter-atomic separation is less than a
wavelength. A remarkable demonstration of such two photon
cooperative effects is given in a recent work \cite{13} using the
methods of single molecule spectroscopy.
 Similar results apply to the
case of two photon emission by identical atoms (Fig.1(b)) if the
photons of frequencies $\omega_a$ and $\omega_b$ are emitted
\begin{equation}
\omega_a+\omega_b=2\omega_{0}.
\end{equation}
In this paper we examine such two photon emission processes in a
cavity. It is advantageous to use a cavity for the study of such a
fundamental process as one would not be constrained by the
requirement of small inter-atomic separation. We demonstrate how
high quality cavities can lead to a large two photon Rabi
oscillation involving two atoms. Note that vacuum Rabi
oscillations in the context of a single atom interacting strongly
with vacuum inside a single mode resonant cavity have been studied
extensively \cite{1,2,3,4}. We also note that the two photon
micromaser in a single mode cavity has been realized \cite{8}. In
this paper we consider two different cases of two photon vacuum
Rabi oscillations, $(i)$ two identical atoms interacting with
vacuum in a two mode cavity, $(ii)$ two nonidentical atoms in a
single mode cavity.

 The paper is
organized as follow. In Sec.II we consider the case of two
identical atoms interacting with two modes of a cavity and discuss
 two photon vacuum Rabi oscillations when the photons are
emitted in different modes under resonance condition. In Sec.III
we consider the case of two nonidentical atoms interacting with a
single mode of the cavity. We present both approximate and
analytical results. In Sec.IV, we confirm that the two photon
vacuum Rabi oscillations survive in the limit of small damping in
a high quality cavity. Finally in Sec.V, we present conclusions
and future outlook.
\section{Two Identical Atoms interacting with Vacuum in a Bimodal Cavity}
We consider two identical two level atoms, with transition
frequency $\omega_0$, interacting with two modes of the vacuum
having frequencies $\omega_a$ and $\omega_b$ in a cavity as shown
in Fig.\ref{fig1}(b). The Hamiltonian for the system is
\begin{eqnarray}
H&=&\hbar\omega_a a^{\dag}a+\hbar\omega_bb^{\dag}b+\nonumber\\
&~&\sum_{i=1,2}\hbar\left[\frac{\omega_0}{2}(|e_i\rangle\langle
e_i|-|g_i\rangle\langle g_i|)
+|e_i\rangle\langle g_i|(g_1a+g_2 b)+|g_i\rangle\langle e_i|(g_1a^{\dag}+g_2
b^{\dag})\right],
\label{h1}
\end{eqnarray}
where $a$ and $a^{\dag}$ $(b$ and $b^{\dag})$ are annihilation and
creation operators for first(second) mode of the cavity, $g_1$ and
$g_2$ are the coupling constants. In a frame rotating with
frequency $\omega_0$, the Hamiltonian $(\ref{h1})$ becomes
\begin{eqnarray}
\label{hcase1}
 H&=&-\hbar\Delta a^{\dag}a-\hbar\delta
b^{\dag}b+\sum_{i=1,2} \hbar\left[|e_i\rangle\langle g_i|(g_1a+g_2
b)+|g_i\rangle\langle e_i
|(g_1a^{\dag}+g_2 b^{\dag}) \right],\\
&&\Delta=\omega_0-\omega_a,~~\delta=\omega_0-\omega_b. \nonumber
\end{eqnarray}
We consider the special case of two photon emission {\it i.e.} the
case when the initial state of the atom-cavity system is
\begin{equation}
|\psi(0)\rangle=|e_1,e_2,0,0\rangle.
\end{equation}
Considering all possible states of the system in evolution, the
state of the system at time $t$ can be written as
\begin{eqnarray}
|\psi(t)\rangle&=&c_1(t)|e_1,e_2,0,0\rangle+\frac{1}{\sqrt{2}}
\left(|e_1,g_2\rangle+|g_1,e_2\rangle\right)
\{c_2(t)|1,0\rangle+c_3(t)|0,1\rangle\}\nonumber\\
&&+|g_1,g_2\rangle\{c_{4}(t)|1,1\rangle+c_5(t)|2,0\rangle
+c_6(t)|0,2\rangle\}.
\label{eq9}
\end{eqnarray}
Different terms in the wave function $(\ref{eq9})$ correspond to no photon emission,
 one photon
emission and two photon emission. The photon emission can take place in either
mode. A very interesting aspect of the state $(\ref{eq9})$ is its entangled
nature. This provides a method of producing entangled states, say, entanglement
of two cavity modes \cite{14}.
The time dependent amplitudes $c_i(t)$ are determined by
\begin{eqnarray}
\dot{c_1}&=&-ig_1\sqrt{2}c_2-ig_2\sqrt{2}c_3\nonumber\\
\dot{c_2}&=&i\Delta c_2-ig_1\sqrt{2}c_1-ig_2\sqrt{2}c_4-2ig_1c_5\nonumber\\
\dot{c_3}&=&i\delta c_3-ig_2\sqrt{2}c_1-ig_1\sqrt{2}c_4-2ig_2c_6\nonumber\\
\dot{c_4}&=&i(\Delta+\delta)c_{4}-ig_2\sqrt{2}c_2-ig_1\sqrt{2}c_3\nonumber\\
\dot{c_5}&=&2i\Delta c_{5}-2ig_1c_{2}\nonumber\\
\dot{c_{6}}&=&2i\delta c_{6}-2ig_2c_{3}.
\label{deq}
\end{eqnarray}
The complete solution of Eq.(\ref{deq}) has six eigenvalues
corresponding to those there will be fifteen peak spectrum.
 In order
to understand the nature of the two atom two photon resonance we
present numerical as well as approximate analysis which can
capture the physics of the cooperative process. We consider the
case when detunings to the cavity field are much larger than the
couplings {\it i.e.} $|\Delta|~,~|\delta|
>>g_1~,~g_2$ but $|\Delta+\delta|$ is small, the condition for two
photon
 resonance is $\Delta+\delta= 0$. In such a  case cooperative two photon process should
dominate and single photon processes would be insignificant.
\begin{figure}[h]
\begin{center}
\includegraphics[width=3.5in,height=3.5in]{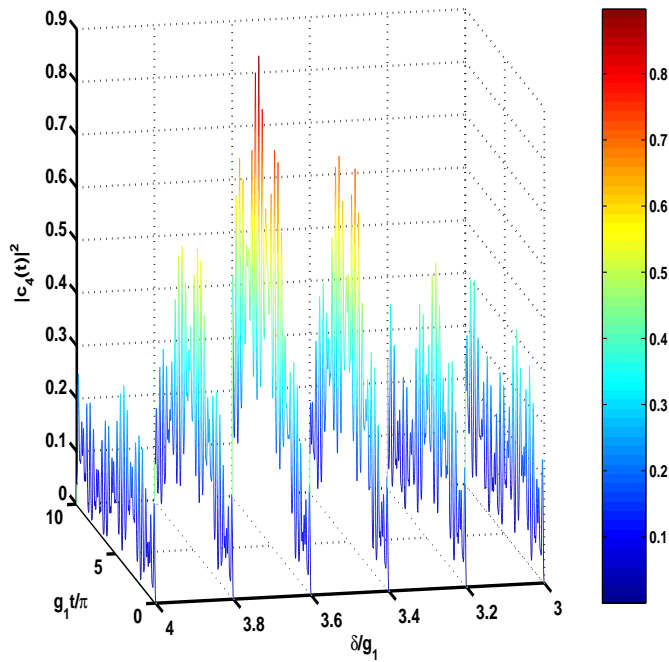}
\end{center}
\caption{(Color online) Two atom two photon emission probability,
$|c_4(t)|^2$ in a system of identical atoms interacting with
vacuum in a two mode cavity, for $g_2/g_1=1.5$ and
$\Delta/g_1=-5.0$.} \label{fig2}
\end{figure}
The results of numerical integration of Eq.(\ref{deq}) are plotted
in Fig.\ref{fig2}. In the case when $g_1\neq g_2$ a novel
resonance is achieved. The probability of two photon emission at
resonance is quite high. The resonance is shifted from the
position $\Delta+\delta=0$. This shift is due to the strong
coupling to the vacuum field in the cavity. For $g_2/g_1=1.5$ and
$\Delta=-5g_1$ maximum two photon emission probability is
approximately $0.9$ and the interaction time required for
achieving maximum probability is given by $g_1t\approx 6\pi$.

Having established numerically that the two photon resonance can
be large in cavities, we present approximate analysis to
demonstrate it. Under the above mentioned conditions for two
photon resonance we can eliminate fast oscillating variables $c_2,
c_3, c_5$, $c_6$ and effectively reduce the dynamics in terms of
slowly oscillating variables $c_1$ and $c_4$. A simple treatment
where one sets $\dot{c}_2=\dot{c}_3=\dot{c}_5=\dot{c}_6=0$ does
not yield the physics of the two atom two photon emission. We thus
relegate the procedure for eliminating fast variables to the
appendix. The reduced form of the Eq.(\ref{deq}) is written as
\begin{eqnarray}
\dot{c_1}&=&-i\left(\frac{2g_1^2\Delta}{\Delta^2-2g_1^2}+\frac{2g_2^2\delta}{\delta^2-2g_2^2}\right)c_1
+2ig_1g_2\left(\frac{\Delta}{\Delta^2+2g_1^2}+\frac{\delta}{\delta^2+2g_2^2}\right)c_4\nonumber\\
\dot{c_4}&=&2ig_1g_2\left(\frac{\Delta}{\Delta^2+2g_1^2}+\frac{\delta}{\delta^2+2g_2^2}\right)c_1
+i\left(\Delta+\delta-\frac{2g_1^2\delta}{\delta^2-2g_2^2}-\frac{2g_2^2\Delta}{\Delta^2-2g_1^2}\right)c_4~.
\label{rdeq}
\end{eqnarray}
The solution of Eq.(\ref{rdeq}) gives
\begin{eqnarray}
\label{result} |c_{4}(t)|^2&=&\frac{4G^2}{4G^2+\Omega^2}
\sin^2\frac{\sqrt{4G^2+\Omega^2}t}{2},\\ {\rm with}~~G&=&
2g_1g_2\left(\frac{\Delta}{\Delta^2+2g_1^2}+\frac{\delta}{\delta^2+2g_2^2}\right)~,~
\Omega=\Delta+\delta+2(g_1^2-g_2^2)\left(\frac{\Delta}{\Delta^2-2g_1^2}-\frac{\delta}{\delta^2-2g_2^2}\right)
\label{omega}
\end{eqnarray}
Note that in the limit $g_1=g_2$ and $\Delta+\delta= 0$, the
probability amplitude $c_4$ for two photon emission tends to zero,
as both $\Omega$ and the numerator in Eq.$(\ref{result})$ become
proportional to $(\Delta+\delta)$. Thus when couplings to the
modes are same two photon emission probability has no resonance.
In this case the transitions from $|e_1,e_2,0,0\rangle$ to
$|g_1,g_2,1,1\rangle$ via states
$\frac{1}{\sqrt{2}}(|e_1,g_2\rangle+|g_1,e_2\rangle)|1,0\rangle$
and $\frac{1}{\sqrt{2}}
(|g_1,e_2\rangle+|e_1,g_2\rangle)|0,1\rangle$ interfere
destructively. We further note that to order $g_1^2g_2^2$ the two
photon resonance does not occur
\begin{equation}
|c_4(t)|^2=\frac{16g_1^2g_2^2}{\delta^2\Delta^2}\sin^2\frac{\delta
t}{2}\sin^2\frac{\Delta t}{2}.
\end{equation}
The usual second order perturbation theory cannot lead to
inter-atomic two photon resonance. One has to consider
\textbf{higher order terms in $g_1$ and $g_2$}. However then the
excitation itself would be negligible. Therefore one needs high
quality cavities.
\begin{figure}[h]
\begin{center}
\includegraphics[width=3.5in,height=3in]{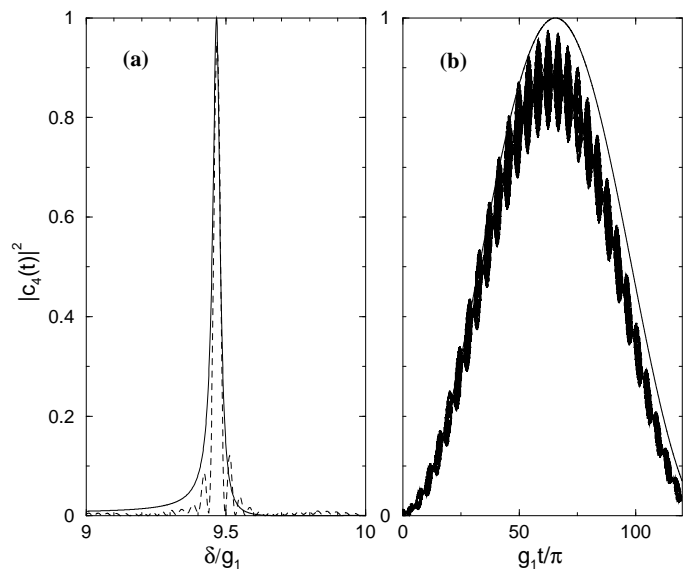}
\end{center}
\caption{The maximum value of the two atom two photon emission
probability, $|c_4(t)|^2$  in the system of two identical atoms
interacting with vacuum in a two mode cavity, is plotted with
respect to (a) detuning $\delta$ and (b) time, for $g_2/g_1=1.5$
and $\Delta=-10 g_1$.  The solid line is corresponding to
approximate result and the dotted line (...)  corresponding to
exact numerical result.} \label{fig3}
\end{figure}
The probability of cooperative emission of two photons in
different modes is a periodic function of time. In
Fig.\ref{fig3}(a), we plot the maximum value of $|c_4(t)|^2$ as a
function of $\delta$ and in Fig.\ref{fig3}(b) as a function of
time $t$, for fixed values of $g_1,g_2$ and $\Delta$.
 At two photon resonance the probability corresponding to two photon emission in one
of the two modes is much smaller than the probability of two
photon emission in different modes. From Eqs. $(\ref{result})$ and
$(\ref{omega})$ it is clear that the two photon resonance occurs
at $\Delta+\delta+4\left(g_1^2/\Delta+g_2^2/\delta\right)
\approx0$. Thus the interaction with the cavity modifies the
condition of two photon resonance. This is seen quite clearly in
Fig.\ref{fig3}(a). We note the connection of the resonance
frequency $\Omega$ to the one photon Stark shifts. It is well
known that the shift in the frequency of a two level atom in the
presence of a field with $n$ photons is given by
$2g^2(n+1)/\Delta$ which is equal to $4g^2/\Delta$ for $n=1$. Thus
the change $4\left(g_1^2/\Delta+g_2^2/\delta\right)$ is equal to
the frequency shift of both the atoms due to the presence of a
single photon. We have checked using the full solution of the
Schrodinger equation that the result $(\ref{result})$ is quite
good. However it should be borne in mind that the exact result is
not periodic and exhibits rapid variations though the envelop
agrees with the result (\ref{result}). The above mentioned
approximate results are valid for larger values of detunings but
for larger values of detutings a large interaction time is
required to reach the maximum of two atom two photon transition
probability. This should be possible with the recently developed
method of trapping atoms in a cavity \cite{kimble}. The other
possibility is to work under the conditions of the Fig.\ref{fig2}.

\section{Two photon emission by two nonidentical Atoms in a single mode cavity}
\begin{figure}[h]
\begin{center}
\includegraphics[width=3.5in,height=3.5in]{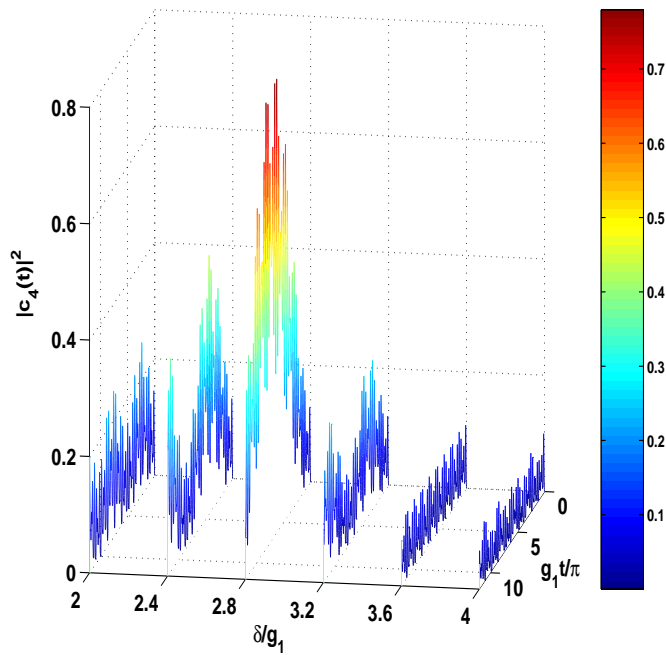}
\end{center}
\caption{(Color online) Two atom two photon emission  probability,
$|c_4(t)|^2$ in a system
 of nonidentical atoms interacting with vacuum in a single mode cavity, for
$\Delta/g_1=-5.0$ and $g_2/g_1=2.0$.} \label{fig4}
\end{figure}

In this section we analyze a system of two nonidentical atoms
interacting with a single mode vacuum field in a cavity
(Fig.\ref{fig1}(a)). Consider two nonidentical two level atoms
having their excited states $|e_1\rangle, ~|e_2\rangle$ and their
ground states $|g_1\rangle, ~|g_2\rangle$ interacting with a
single mode cavity-field of frequency $\omega$. The Hamiltonian of
this system is
\begin{eqnarray}
H&=&\hbar\left[\frac{\omega_1}{2} (|e_1\rangle\langle e_1|-|g_1\rangle\langle g_1|)
+\frac{\omega_2}{2}(|e_2\rangle\langle e_2|-|g_2\rangle\langle g_2|)
+\omega a^{\dag}a\right]\nonumber\\
&~&+\hbar g_1\left(|e_1\rangle\langle g_1|a+a^{\dag}
|g_1\rangle\langle e_1|\right)+\hbar g_2\left(|e_2\rangle\langle
g_2|a+a^{\dag}|g_2\rangle\langle e_2|\right),
\end{eqnarray}
where $\omega_1 (\omega_2)$ is transition frequency for first
(second) atom,
 $a$ and $a^{\dag}$ are annihilation and creation
operators for the field, and $g_1(g_2)$ is the coupling constant
to the cavity mode with first(second) atom . In a rotating frame
the Hamiltonian $H$ can be written as
\begin{eqnarray}
H&=&-\hbar\Delta|g_1\rangle\langle g_1|
 -\hbar\delta|g_2\rangle\langle
 g_2|+\hbar g_1\left(|e_1\rangle\langle g_1|a+a^{\dag}
|g_1\rangle\langle e_1|\right)+\hbar g_2\left(|e_2\rangle\langle
g_2|a+a^{\dag}|g_2\rangle\langle e_2|\right),\\
&&\Delta = \omega_1-\omega, \delta=\omega_2-\omega.
\end{eqnarray}
Let us consider an initial state $|\psi(0)\rangle=|e_1,e_2,0\rangle$ with both
atoms in the excited state and cavity in the vacuum state.
The state of the system at time $t$ can be written as
\begin{equation}
|\psi(t)\rangle=
c_1(t)|e_1,e_2,0\rangle+c_2(t)|e_1,g_2,1\rangle+c_3(t)|g_1,e_2,1\rangle
+c_4(t)|g_1,g_2,2\rangle,
\end{equation}
where the expansion coefficients c's satisfy
\begin{eqnarray}
\dot{c_1}&=&-ig_2c_2-ig_1c_3,\nonumber\\
\dot{c_2}&=&i\delta c_2-ig_1\sqrt{2}c_4-ig_2c_1,\nonumber\\
\dot{c_3}&=&i\Delta c_3-ig_2\sqrt{2}c_4-ig_1c_1,\nonumber\\
\dot{c_4}&=&i\left(\Delta+\delta\right)c_4-ig_1\sqrt{2}c_2-ig_2\sqrt{2}c_3.
\label{case2}
\end{eqnarray}
The two photon resonance condition for this system would be
$\Delta+\delta=0$. For couplings $g_1~,~g_2$, much smaller than
$|\Delta|,~|\delta|$, the solution of Eq.(\ref{case2}) gives
\begin{eqnarray}
c_4(t)&=&-\frac{4g_1g_2\sqrt{2}}{\delta\Delta}\sin\frac{\delta
t}{2}\sin\frac{\Delta t}{2}+{\rm higher~~order~~terms}.
\label{mode2}
\end{eqnarray}
The first term in Eq.(\ref{mode2}) represents independent emission
by each atom. Clearly, to lowest order in $g_1g_2$ no two photon
resonance occurs. Such a resonance can come from the terms of the
higher order. Assuming that $|\Delta|$ and $|\delta|$ are large
but $|\Delta+\delta|$ is small, we eliminate fast oscillating
variables $c_2$ and $c_3$ in a way similar to the previous case
and the Eq.(\ref{case2}), in terms of slowly oscillating variables
reduces, to
\begin{eqnarray}
\dot{c_1}&=&-i\left(\frac{g_1^2}{\Delta}+\frac{g_2^2}{\delta}\right)c_1
+ig_1g_2\sqrt{2}\left(\frac{\Delta}{\Delta+2g_1^2}+\frac{\delta}{\delta+2g_2^2}\right)c_4,\nonumber\\
\dot{c_4}&=&ig_1g_2\sqrt{2}\left(\frac{\Delta}{\Delta+2g_1^2}+\frac{\delta}{\delta+2g_2^2}\right)c_1
+i\left(\Delta+\delta+\frac{2g_1^2}{\Delta}+\frac{2g_2^2}{\delta}\right)c_4.
\label{rcase2}
\end{eqnarray}
 We find the approximate result for the
two photon emission probability
\begin{eqnarray}
\label{result2} |c_{4}(t)|^2&=&\frac{4G'^2}{4G'^2+\Omega'^2}
\sin^2\frac{\sqrt{4G'^2+\Omega'^2}t}{2},\\ {\rm with}~~G'&=&
\sqrt{2}g_1g_2\left(\frac{\Delta}{\Delta^2+2g_1^2}+\frac{\delta}{\delta^2+2g_2^2}\right)~~,
\Omega'=\Delta+\delta+3\left(\frac{g_1^2}{\Delta}+\frac{g_2^2}{\delta}\right).
\label{omega2}
\end{eqnarray}
For large $|\Delta|$ and $|\delta|$ the Eq.(\ref{result2}) shows
two photon resonance at
$\Delta+\delta+3(g_1^2/\Delta+g_2^2/\delta)\approx0$. Further such
two atom two photon resonance appears for $g_1\neq g_2$, which
disappears when $g_1=g_2$. In the latter case the antisymmetric
state $\left(|g_1,e_2,1\rangle-|e_1,g_2,1\rangle\right)/\sqrt{2}$
is decoupled from $|e_1,e_2,0\rangle$ and $|g_1,g_2,2\rangle$. We
present numerical results in Fig.\ref{fig4} . The graph shows two
photon resonance for $g_1\neq g_2$. It is clear that the position
of resonance is shifted from $\Delta+\delta=0$. This shift in the
position of resonance is due to larger values of $g_1$ and $g_2$,
and depends on the ratio $g_2/g_1$. There is a large enhancement
in the probability of two photon resonant emission in a high
quality cavity. It is expected that such effects can be studied by
placing the system used by Hettich {\it et al}\cite{13} in a
cavity.
\section{effects of cavity damping}
\begin{figure}[h]
\begin{center}
\includegraphics[width=5in,height=3in]{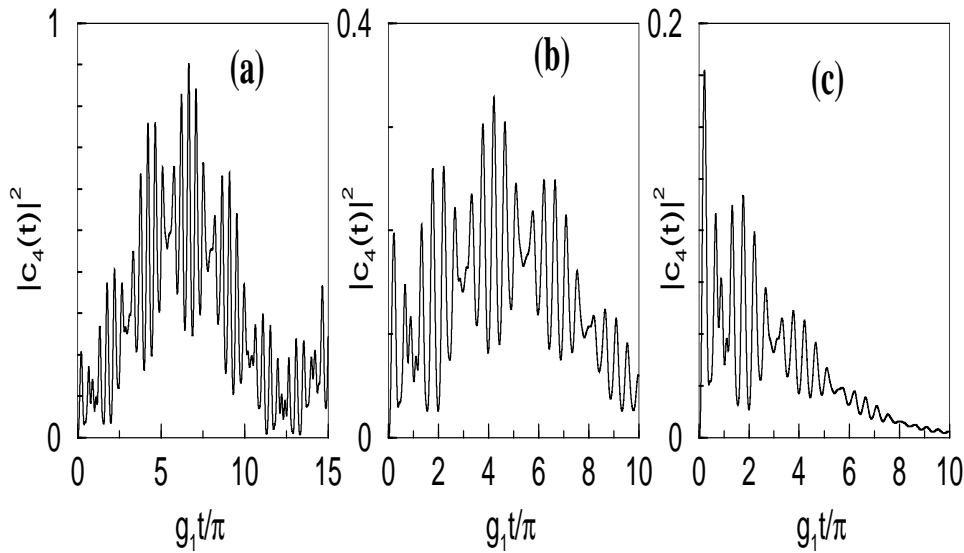}
\end{center}
\caption{Periodic behavior of two atom two photon emission
probability $|c_4(t)|^2$, for identical atoms interacting with
vacuum in a bimodal cavity, for $\delta=3.5g_1$, $\Delta=-5g_1$,
$g_2=1.5g_1$ and  cavity damping constants $(a)$
$\kappa_a=\kappa_b=0.00$, $(b)$ $\kappa_a=\kappa_b=0.03g_1$, $(c)$
$\kappa_a=\kappa_b=0.1g_1$.} \label{fig5}
\end{figure}

Before concluding we examine the effect of cavity decay on two
atom two photon vacuum Rabi oscillations. We do a calculation
based on master equation. Let $2\kappa_a$ and $2\kappa_b$ be the
rate of loss of photons from the first mode and the second mode
respectively. The density matrix of the system of two atoms
interacting with two mode field in the cavity will evolve
according to the master equation
\begin{equation}
\dot{\rho}=-\frac{i}{\hbar}[H,\rho]-\kappa_a\left(a^{\dag}a\rho-2a\rho a^{\dag}
+\rho a^{\dag}a\right)-\kappa_b\left(b^{\dag}b\rho-2b\rho b^{\dag}
+\rho b^{\dag}b\right).
\label{meq}
\end{equation}
The density matrix for this system can be expressed in terms of all the states
which are generated by the combined effect of $H$ and dissipation.
Because of the cavity decay, many more states are involved in the dynamics.
 For example
for identical atoms interacting in a bimodal cavity, the relevant states are
$|e_1,e_2,0,0\rangle$, $|g_1,e_2,0,0\rangle$, $|g_1,e_2,1,0\rangle$,
$|g_1,e_2,0,1\rangle$, $|e_1,g_2,0,0\rangle$, $|e_1,g_2,1,0\rangle$,
$|e_1,g_2,0,1\rangle$, $|g_1,g_2,0,0\rangle$, $|g_1,g_2,0,1\rangle$,
$|g_1,g_2,1,0\rangle$, $|g_1,g_2,0,2\rangle$, $|g_1,g_2,1,1\rangle$, and
$|g_1,g_2,2,0\rangle$. For this system density matrix is expressed as
\begin{equation}
\rho\equiv\sum_{i',j',i,j=0}^{1}\sum_{k'=0}^{i'+j'}\sum_{k=0}^{i+j}\sum_{l'=0}^{
i'+j'-k'}\sum_{l=0}^{i+j-k}\rho(i',j',k',l',i,j,k,l)|i',j',k',l'\rangle\langle
i,j,k,l|~~.
\label{dmatrix}
\end{equation}
Here $i, i'$ $(j, j')$ represent states of the first (second) atom
with the convention $|0\rangle$ corresponding to excited state and
$|1\rangle$ corresponding to ground state, the indices $k,~k'$
$(l,~l')$ represent the number of photons in the first (second)
mode. Thus the dissipation requires considerable numerical work.
Results for two identical atoms in a bimodal cavity are shown in
Fig.\ref{fig5}. We show  results for optical cavities with
$g/\kappa\approx 30$ in Fig.\ref{fig5}(b) and  for currently
realizable cavities $(g/\kappa=10)$ in Fig.\ref{fig5}(c). The two
atom two photon vacuum Rabi oscillations survive in the limit of
small damping $g/\kappa\approx 30$ but for larger damping
$(g/\kappa=10)$ die fast. Similar results are found for two
nonidentical atoms in a single mode cavity.
\section{conclusions}
We have reported large two atom two photon vacuum Rabi
oscillations in two systems, one having two identical atoms in a
two-mode cavity and another having two nonidentical atoms in a
single-mode cavity. We have shown that for asymmetric couplings
$(g_1\neq g_2)$, the probability of two photon emission is quite
large but for symmetric couplings $(g_1=g_2)$, the two photon
emission probability is very small. Further, we have shown that
the condition of two photon resonance in the case of strong
atom-field interaction is modified from its free-space form
$(\Delta+\delta=0)$. These two photon transitions involving two
atoms can be used for generating and detecting different types of
entanglement between two field modes and two atoms \cite{15}.

GSA thanks V. Sandoghdar and G. Rempe for discussions on the two
atom two photon resonance.

\appendix
\section{}
Our procedure for eliminating fast oscillating variables is
extended form of the procedure discussed in Ref.\cite{adb}. The
Hamiltonian (\ref{hcase1}) can be written as
\begin{eqnarray}
&&H=H_0+\epsilon V,\nonumber\\
&&{\rm where}~~\nonumber\\&&H_0=-\Delta a^{\dag}a-\delta
b^{\dag}b,~~\epsilon V=\sum_{i=1,2} \hbar\left[|e_i\rangle\langle
g_i|(g_1a+g_2 b)+|g_i\rangle\langle e_i |(g_1a^{\dag}+g_2
b^{\dag}) \right].
\end{eqnarray}
The eigenstates and corresponding eigenvalues of $H_0$ are
\begin{eqnarray}
&&|1\rangle\equiv|e_1,e_2,0,0\rangle~~~~~~~~~~~~~~~~~~~~~~~~~~~~~~~~~E_1\equiv0\nonumber,\\
&&|2\rangle\equiv2^{-1/2}(|e_1,g_2\rangle+|g_1,e_2\rangle)|1,0\rangle~~~~~~~~~~E_2\equiv-\Delta\nonumber,\\
&&|3\rangle\equiv2^{-1/2}(|e_1,g_2\rangle+|g_1,e_2\rangle)|0,1\rangle~~~~~~~~~~E_3\equiv-\delta\nonumber,\\
&&|4\rangle\equiv|g_1,g_2,1,1\rangle~~~~~~~~~~~~~~~~~~~~~~~~~~~~~~~~~E_4\equiv-(\Delta+\delta)\nonumber,\\
&&|5\rangle\equiv|g_1,g_2,2,0\rangle~~~~~~~~~~~~~~~~~~~~~~~~~~~~~~~~~E_5\equiv-2\Delta\nonumber,\\
&&|6\rangle\equiv|g_1,g_2,0,2\rangle~~~~~~~~~~~~~~~~~~~~~~~~~~~~~~~~~E_6\equiv-2\delta\nonumber.
\end{eqnarray}
The resolvent for $H_0$ is the function
\begin{equation}
G_0(z)=\frac{1}{z-H_0},
\end{equation}
where $z$ is complex. If $P_i$ is projection operator for the
eigenstates of $H_0$
\begin{equation}
P_i=|i\rangle\langle i|;~~i=1,2...6.
\end{equation}
The resolvent $G_0$ can be expressed as
\begin{equation}
G_0(z)=\sum_i\frac{P_i}{z-E_i}.
\end{equation}
The resolvent for the full Hamiltonian $H$ is
\begin{eqnarray}
G(z)&=&\frac{1}{z-H_0-\epsilon V},\nonumber\\
&=&\frac{1}{z-H_0}\left(1+\epsilon
V\frac{1}{z-H}\right),\nonumber\\
\label{resol}
&=&G_0(1+\epsilon V G).
\end{eqnarray}
From Eq.(\ref{resol}) the resolvent for the full Hamiltonian $H$
can be expressed in the power series of $\epsilon$ as
\begin{equation}
G=\sum_n \epsilon^n G_0(V G_0)^n.
\label{expan}
\end{equation}
\begin{figure}[h]
\begin{center}
\includegraphics[width=4in,height=1in]{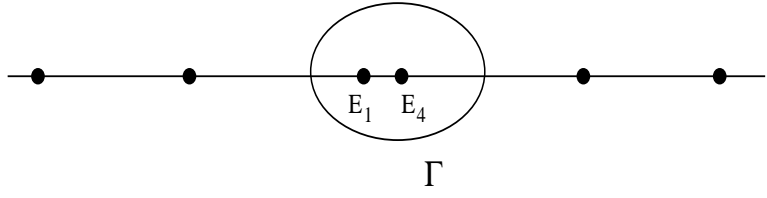}
\end{center}
\caption{ The contour in complex plane shielding two eigenvalues
$E_1$ and $E_4$ and leaving others outside.} \label{fig6}
\end{figure}
For small values of $\epsilon$, $G(z)$ has singularities in the
complex z-plane in the neighborhood of poles of function $G_0$
{\it i.e.} eigenvalues of $H_0$. Further eigenvalues $E_1$ and
$E_4$ are very close to each other under the condition
$\Delta+\delta\approx0$ and other eigenvalues are largely
separated. We consider a contour, $\Gamma$ in the z-plane that
encloses eigenvalues $E_1$ and $E_4$ only and leaves others
outside as shown in the Fig.\ref{fig6}. We define a new projection
operator $P_{\Gamma}$ as
\begin{eqnarray}
P_{\Gamma}&=&\bar{P}_1+\bar{P}_4, \nonumber\\
&=&\frac{1}{2i\pi}\oint_{\Gamma}G(z)dz.
\end{eqnarray}
Here $\bar{P}_1$ and $\bar{P}_4$ are the projection operators for
eigenstates of full Hamiltonian $H$ corresponding to the
eigenvalues inside the contour. The effective Hamiltonian will
have the form
\begin{equation}
H_{eff}\equiv (P_1+P_4)HP_{\Gamma}(P_1+P_4).
\end{equation}
From the definition of the resolvent we have
\begin{eqnarray}
&&(z-H)G\equiv G(z-H)\equiv1.\nonumber\\
\label{hproj}
 &&HP_{\Gamma}=\frac{1}{2i\pi}\oint_{\Gamma}z G(z)dz.
\end{eqnarray}
Substituting value of $G(z)$ from Eq.(\ref{expan}) in
Eq.(\ref{hproj}) and interchanging summation to the integration we
have
\begin{eqnarray}
H P_{\Gamma}=\sum_n\frac{1}{2i\pi}\oint_{\Gamma}z G_0(VG_0)^n dz.
\end{eqnarray}
The effective Hamiltonian can be expressed as
\begin{eqnarray}
H_{eff}&=&E_1P_1+E_4P_4+\sum_{n=1}^{\infty}\epsilon^n
A^{(n)};\nonumber\\
\label{nth}
A^{(n)}&=&(P_1+P_4)\sum_{n=1}^{\infty}\frac{1}{2i\pi}\oint_{\Gamma}z
G_0(VG_0)^n dz (P_1+P_4).
\end{eqnarray}
Inside the contour $\Gamma$, $G_0$ has singularities at $E_1$ and
$E_4$ only so the integral in the Eq.(\ref{nth}) is nothing but
the sum of the residues at $z=E_1$ and $z=E_4$. Further as in our
case $\epsilon P_1VP_1$, $\epsilon P_4VP_4$ and $\epsilon P_1VP_4$
equal to zero, there is no first order and third order terms. The
second order term is
\begin{eqnarray}
&&A^{(2)}=P_1VQ_1VP_1+P_4VQ_4VP_4+P_1VQ_4VP_4+P_4VQ_4VP_1~;\\
&&Q_j=\sum_{i\neq1,4}\frac{P_i}{E_j-E_i}.\nonumber
\end{eqnarray}
The forth order term is
\begin{eqnarray}
A^{(4)}&=&\frac{1}{2i\pi}\oint_{\Gamma}z\left(\frac{P_1}{z-E_1}
+\frac{P_4}{z-E_4}\right)V\sum_{i\neq1,4}\frac{P_i}{z-E_i}V\left(\frac{P_1}{z-E_1}
+\frac{P_4}{z-E_4}+\sum_{j\neq1,4}\frac{P_j}{z-E_j}\right)V\nonumber\\
&&\sum_{k\neq1,4}\frac{P_k}{z-E_k} V\left(\frac{P_1}{z-E_1}
+\frac{P_4}{z-E_4}\right)dz
\end{eqnarray}
For simplification we use the condition for resonance
$\Delta+\delta=0$, {\it i.e.} $E_1=E_4$. Thus the forth order term
is
\begin{eqnarray}
A^{(4)}=\frac{1}{2i\pi}\oint_{\Gamma}z\left(\frac{P_1}{z-E_1}
+\frac{P_4}{z-E_1}\right)V\sum_{i\neq1,4}\frac{P_i}{z-E_i}V\sum_{j\neq1,4}\frac{P_j}{z-E_j}V
\sum_{k\neq1,4}\frac{P_k}{z-E_k} V\left(\frac{P_1}{z-E_1}
+\frac{P_4}{z-E_1}\right)dz. \label{forth}
\end{eqnarray}
Integrating Eq.(\ref{forth}) we have the forth order term
\begin{equation}
A^{(4)}=\left(P_1+P_4\right)VQ_1VQ_1VQ_1V(P_1+P_4).
\end{equation}
Using the values of $E_1,~E_2,~E_3,~E_4,~E_5,~E_6$ and $V$ the
effective Hamiltonian expressed in basis $|e_1,e_2,0,0\rangle$ and
$|g_1,g_2,1,1\rangle$ is
\begin{eqnarray}
H_{eff}=\left[\begin{array}{cc}
\frac{2g_1^2}{\Delta}+\frac{2g_2^2}{\delta}+\frac{4g_1^4}{\Delta^3}+\frac{4g_2^4}{\delta^3}&
-\frac{2g_1g_2}{\Delta}-\frac{2g_1g_2}{\delta}+\frac{4g_1^3g_2}{\Delta^3}+\frac{4g_1g_2^3}{\delta^3}\\
-\frac{2g_1g_2}{\Delta}-\frac{2g_1g_2}{\delta}+\frac{4g_1^3g_2}{\Delta^3}+\frac{4g_1g_2^3}{\delta^3}&
-(\Delta+\delta)-\frac{2g_2^2}{\delta}-\frac{2g_1^2}{\Delta}+\frac{4g_1^2g_2^2}{\Delta^3}
+\frac{4g_1^2g_2^2}{\delta^3}
\end{array}\right].
\end{eqnarray}
With some algebraic manipulation and considering $g_1$ and $g_2$
up to forth order effectively the Hamiltonian (\ref{hcase1})
reduces to
\begin{eqnarray}
H_{eff}=\left[\begin{array}{cc}
\frac{2g_1^2\Delta}{\Delta^2-2g_1^2}+\frac{2g_2^2\delta}{\delta^2-2g_2^2}&
-2g_1g_2\left(\frac{\Delta}{\Delta^2+2g_1^2}+\frac{\delta}{\delta^2+2g_2^2}\right)\\
-2g_1g_2\left(\frac{\Delta}{\Delta^2+2g_1^2}+\frac{\delta}{\delta^2+2g_2^2}\right)&
-\left(\Delta+\delta-\frac{2g_2^2\Delta}{\Delta^2-2g_1^2}-\frac{2g_1^2\delta}{\delta^2-2g_2^2}\right)
\end{array}\right].
\label{effective}
\end{eqnarray}
It should be noted here as two atom two photon resonance appears
at large interaction time in dispersive limit, the terms in the
 effective Hamiltonian up to forth order are important to predict correct
evolution. Using the effective Hamiltonian (\ref{effective}) the
Eq.(\ref{deq})
 reduces to Eq.(\ref{rdeq}).
\end{document}